\documentclass[12pt]{article}

\usepackage[margin=1in]{geometry}

\usepackage[utf8]{inputenc}

\usepackage{amsmath, amssymb, amsfonts}
\usepackage{mathtools}  % 可選加強 math 表達式

\usepackage{graphicx}
\usepackage{float}  % 強化圖表固定位置
\usepackage{caption}
\usepackage{subcaption}

\usepackage[colorlinks=true, linkcolor=blue, citecolor=blue, urlcolor=blue]{hyperref}

\usepackage[numbers,sort&compress]{natbib}  % astro-ph 常用這套

\usepackage{titlesec}
\titleformat{\section}{\large\bfseries}{\thesection}{1em}{}
\titleformat{\subsection}{\normalsize\bfseries}{\thesubsection}{1em}{}

\usepackage{algorithm}
\usepackage{algorithmic}

\usepackage{setspace}
\usepackage{url}

\usepackage{booktabs}
\usepackage{multirow}
\usepackage{array}

\usepackage{xcolor}

\usepackage{authblk}  % 新增這行

\title{An Exploratory Framework for Future SETI Applications: Detecting Generative Reactivity via Language Models}

\author[1,2]{Po-Chieh Yu}

\affil[1]{Taiwan Astronomical Research Alliance (TARA), Taiwan}
\affil[2]{Institute of Astronomy and Astrophysics, Academia Sinica, Taipei, 10617, Taiwan}

\date{}

\begin{document}
\maketitle
\normalsize

\begin{abstract}
We present an exploratory framework to test whether noise-like input can induce structured responses in language models. Instead of assuming that extraterrestrial signals must be decoded, we evaluate whether inputs can trigger linguistic behavior in generative systems. This shifts the focus from decoding to viewing structured output as a sign of underlying regularity in the input. We tested GPT-2 small, a 117M-parameter model trained on English text, using four types of acoustic input: human speech, humpback whale vocalizations, \textit{Phylloscopus trochilus} birdsong, and algorithmically generated white noise. All inputs were treated as noise-like, without any assumed symbolic encoding. To assess reactivity, we defined a composite score called Semantic Induction Potential (SIP), combining entropy, syntax coherence, compression gain, and repetition penalty. Results showed that whale and bird vocalizations had higher SIP scores than white noise, while human speech triggered only moderate responses. This suggests that language models may detect latent structure even in data without conventional semantics. We propose that this approach could complement traditional SETI methods, especially in cases where communicative intent is unknown. Generative reactivity may offer a different way to identify data worth closer attention.
\end{abstract}

\section{Introduction}

Over the past 40 years, the Search for Extraterrestrial Intelligence \citep*[SETI;][]{cocconi1959search,drake1961,shuch2011ozma,Tarter2001SETI} has reported candidate signals, including the “Wow!” signal \citep{kraus1994wow} and more recent anomalies near Proxima Centauri \citep*{Smith2021Proxima}. However, none have been independently confirmed \citep{Gray2001Wow,Gray2002WowFollowup}, and the latest detections have been attributed to terrestrial interference \citep{sheikh2021blc1}.

Conventional SETI approaches typically assume that extraterrestrial signals carry specific and decodable content, such as mathematical sequences and narrowband transmissions that are designed to be received by another civilization. These assumptions imply that extraterrestrial civilizations have advanced technological capability and communicative intent. However, alternative perspectives like the Zoo Hypothesis \citep{Ball1973,Forgan2016,Crawford2024} and Dysonian SETI \citep{Bradbury2011} suggest that extraterrestrial intelligence may either avoid direct contact or express itself through indirect, observable phenomena. These views support using a wider range of methods to interpret anomalous signals.

\citet{wright2018visions} indicated that SETI often reveals more about our conceptual limitations than about extraterrestrial intentions, highlighting the role of self-reflection in signal interpretation. \citet{cabrol2016alien} further emphasizes the need to develop new frameworks that move beyond assumptions tied to human senses and languages.
They proposed that more diverse cognitive and interpretive models should be incorporated into future detection strategies.
Based on these arguments, rather than assuming the data must be decoded, we explore whether certain input might trigger structured responses in generative models. 
Such responses, especially in systems trained on human text, could provide a measurement of detectability in data with latent structure.

Most SETI workflows focus on measurable features, such as narrowband peaks or repeating patterns. Segments that appear broadband, aperiodic, or spectrally flat are typically dismissed as noise. Large portions of SETI data are never examined for hidden structure beyond statistical characteristics. As a result, many of these segments are treated as meaningless, even though some may contain forms of complexity that conventional methods fail to detect.

Rather than asking whether the data is interpretable in human terms, we ask whether it can trigger linguistic behavior in generative models. While recent studies have applied machine learning to SETI signal classification \citep*{chen2022edge, cox2018classification, harp2019machine, Ma2023DLSETI}, the rise of large language models (LLMs) \citep*[e.g.,][]{Radford2019GPT2,brown2020language,mistral2023,openai2023gpt4,touvron2023llama} offers a different approach: generating structured output from inputs that carry no specific meaning and do not follow any known communicative conventions. This kind of model reactivity may reveal latent patterns in data that remain undetectable using conventional methods. Unlike previous work which analyzes potentially meaningful signals for linguistic patterns \citep{elliott2004corpus}, our method begins with unfiltered, noise-like data and observes whether any internal regularity is sufficient to trigger language-like responses.

In this work, we present a framework for testing structural reactivity in LLMs, using input from human language, animal vocalizations, and white noise. These inputs provide a baseline for evaluating model responses to different levels of latent structure. We begin by outlining the conceptual hypothesis and methods (Section 2), followed by a demonstration of model responses and interpretive analysis (Section 3). In Section 4, we discuss the broader implications for SETI and signal detection, and summarize key conclusions in Section 5.

\section{Hypothesis and Method}

\subsection{Hypothesis}

We propose that highly advanced extraterrestrial civilizations may favor low-power or indirect signaling strategies over beacon-like transmissions.
Unlike less advanced civilizations, which might rely on amplitude-based signals to maximize detectability, such emissions are vulnerable to distance-related attenuation and may risk revealing the sender’s location. In contrast, more advanced systems may favor low-amplitude, phase-based signals that resemble background noise but contain structured temporal features. While harder to detect using conventional methods, such approach may serve as a more efficient or cautious mode of interstellar communication.

One motivation for adapting such an approach may be the limitations of language itself. Language depends on shared context and is prone to misinterpretation, especially across differing cognitive architectures or cultural frameworks. Communicating based on syntax or symbols assumes some common foundations for processing information, which may not hold in interstellar communication. Alternatively, more generalizable cognitive indicators, such as semantic priming, pattern recognition, or responses triggered by hidden structure may offer more robust markers of intelligence. In this framing, the input data does not need to be decoded as a message, but can be evaluated for its capacity to trigger structured behavior in a highly advanced system.

\citet{brin2014meti} questions whether contact strategies based on message decoding are realistic, highlighting the unpredictability of how intelligence may manifest. Our proposed signaling strategy addresses this concern by reducing the risk of cultural contamination or unintended technological exchange. It allows contact to emerge only when the receiver is cognitively prepared. In this context, phase-based input data may exhibit internal complexity, energy efficiency, and long-term stability. While inaccessible to systems with limited abstraction capabilities, such patterns may still trigger structured responses in receivers capable of language generation.

Moreover, this framework offers practical advantages in uncertain or noisy environments. Instead of relying on predefined templates, it uses generative reactivity, which makes it less sensitive to interference and unpredictable distortions. For instance, language models often fail to produce structured responses when given unstructured or malformed input. Coherent output is typically generated only when the input contains some form of internal regularity. This provides a kind of semantic fault tolerance, allowing the model to respond more strongly to data that carry latent structure.

\subsection{Method: Probing Structured Output from Noise-Like Data}

We used GPT-2 small \citep{Radford2019GPT2} to test whether noise-like input can trigger structured generative output.
%This shifts the focus of detection away from message interpretation and toward observing how the model responds.
The GPT-2 small is a compact language model trained only on English text, with no exposure to speech or multilingual input. 
Because of its smaller size and narrower training scope, the model is less constrained by human syntax and tends to respond more directly to structural patterns. 

With only 117 million parameters, GPT-2 small is far below the threshold for emergent behaviors \citep{wei2022emergent}, and less prone to hallucination effects typically seen in larger language models. More powerful systems like GPT-3 \citep{brown2020language} and GPT-4 \citep{openai2023gpt4} can generate coherent output even from random input.
However, they are also more likely to hallucinate due to their strong sensitivity to human language priors \citep{ji2023survey}. In contrast, GPT-2 small has a higher threshold for producing structured language. This makes GPT-2 small a conservative testbed for detecting structure in input data, as it reduces the influence of strong linguistic priors that may dominate responses in larger models.

Previous work has explored how large language models respond to minimal prompts, adversarial tokens, and multimodal inputs \citep{wei2022emergent, holtzman2020degeneration}. However, to our knowledge, no framework has used LLMs to detect latent signal structure based on their tendency to generate language-like output.

\subsubsection{Input Types and Sources}

To test whether different input types could induce distinct linguistic responses, we selected four categories of data representing a range of presumed communicative structure: (1) human language (English, containing both meaning and syntax), (2) humpback whale vocalizations (non-human but socially meaningful), (3) Phylloscopus trochilus birdsong (non-semantic but structurally patterned), and (4) white noise (non-communicative and unstructured). All input types were treated as noise-like data. This framing assumes no prior knowledge or semantic labeling, allowing us to evaluate whether structured output can arise purely from the organization present in the input. Each category was segmented into ten 12-second clips. This duration provides sufficient temporal context to trigger structured responses from the language model, while remaining short enough to avoid performance degradation due to excessive prompt length, as noted in previous studies on LLM sensitivity to input size \citep*{levy2024promptlength, liu2023lostmiddle}. The input sources are as follows:

\begin{enumerate}
\item \textbf{Human Language:} We use spoken English data from the LibriVox corpus, which contains public domain recordings of literary works read by volunteers \citep{librivox}. A continuous segment from a single speaker was selected and divided into ten 12-second clips. All LibriVox materials are freely available for research and reuse.

\item \textbf{Whale Vocalizations:} Sound clips of humpback whale vocalizations were downloaded from the SanctSound passive acoustic dataset, provided by the NOAA Office of National Marine Sanctuaries and the U.S. Navy \citep{sanctsound2021}. We used two 60-second recordings collected within the U.S. National Marine Sanctuary System (Sound Clips CI02 and CI05), accessed via \url{https://doi.org/10.25921/saca-sp25}. Due to the limited duration of each recording (60 seconds), we used two separate whale recordings to generate the full set of ten 12-second clips. In contrast to the other categories, which used a single continuous source, this introduces inter-source variation within the whale class. However, it also provides an opportunity to examine how SIP responses vary across naturally diverse vocalizations.

\item \textbf{Birdsong:} Birdsong input was sourced from a \textit{Phylloscopus trochilus} recording available on the Xeno-Canto platform. Specifically, we used XC763801 by \citet{XC763801}, released under the Creative Commons Attribution-NonCommercial-ShareAlike 4.0 license. The recording was segmented into ten 12-second clips. The file is publicly accessible at \url{https://www.xeno-canto.org/763801}.
    
\item \textbf{White Noise:} Synthetic white noise was generated using Python’s NumPy module by sampling from a Gaussian distribution (mean 0, standard deviation 0.1) at a sampling rate of 16~kHz, with a total duration of 120 seconds. The signal was clipped to the range $[-1.0, 1.0]$ and exported as 16-bit signed PCM WAV files using SciPy’s \texttt{write} function. The resulting file was segmented into ten 12-second clips. This procedure follows standard practices in digital signal processing \citep{oppenheim1999signals}.

\end{enumerate}

\subsubsection{Semantic Triggering Detection Pipeline}
We quantify output results using a metric referred to as Semantic Induction Potential (SIP).
The implementation of our procedure, called the Semantic Triggering Detection Pipeline (STDP), is shown in Figure 1 and described below.

\begin{figure}
\centerline{\includegraphics[width=4.0in]{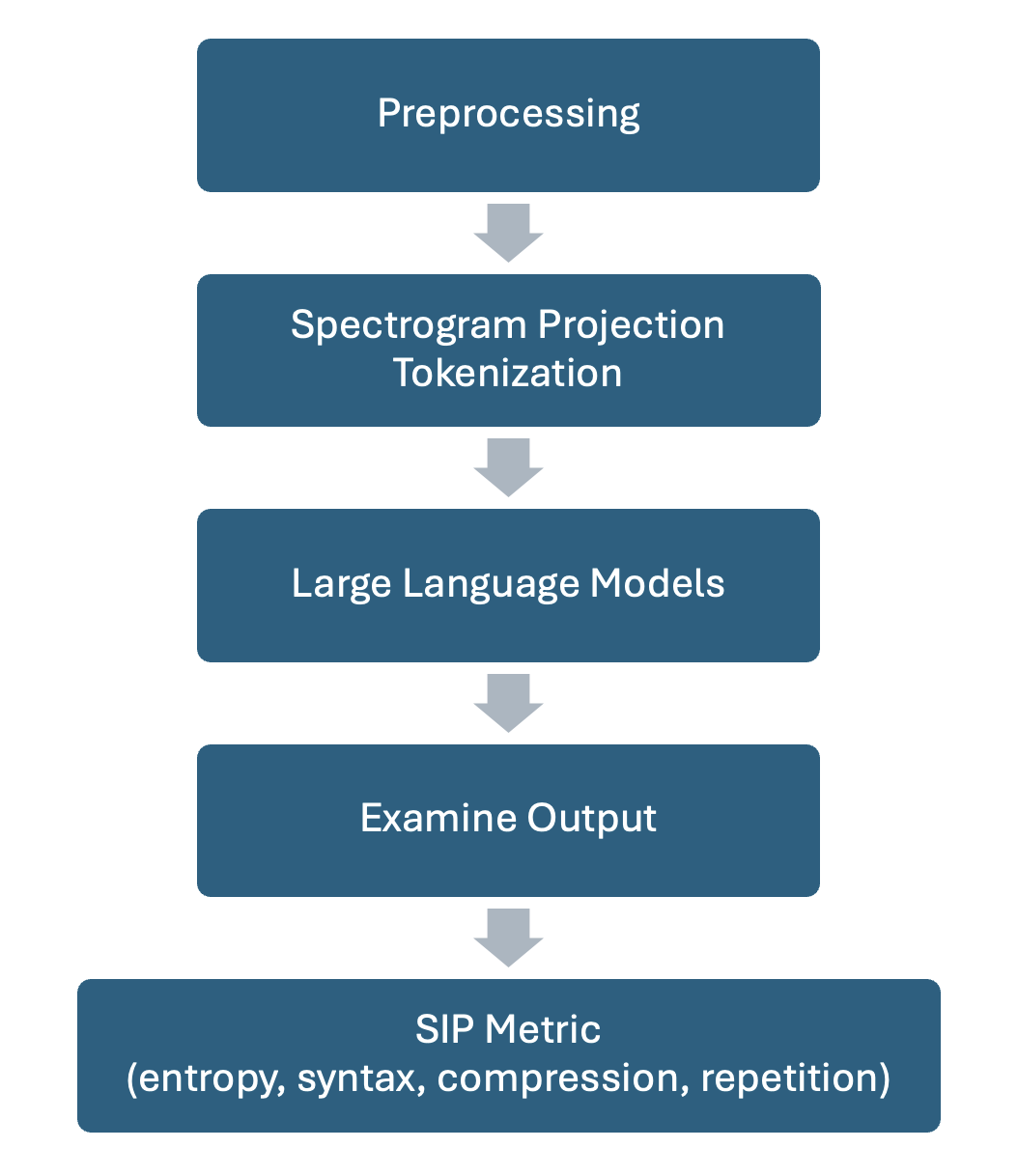}}
%%%call your figure name in the place "figurename.eps"
\caption{Semantic Triggering Detection Pipeline (STDP). Audio inputs are preprocessed and projected into symbolic form via spectral tokenization. These are passed to a language model, and the output is evaluated by the SIP metric, which integrates entropy reduction, syntactic coherence, compression efficiency, and repetition penalty.}
\label{fig_sim}
\end{figure}

\begin{figure}
\centerline{\includegraphics[width=6.5in]{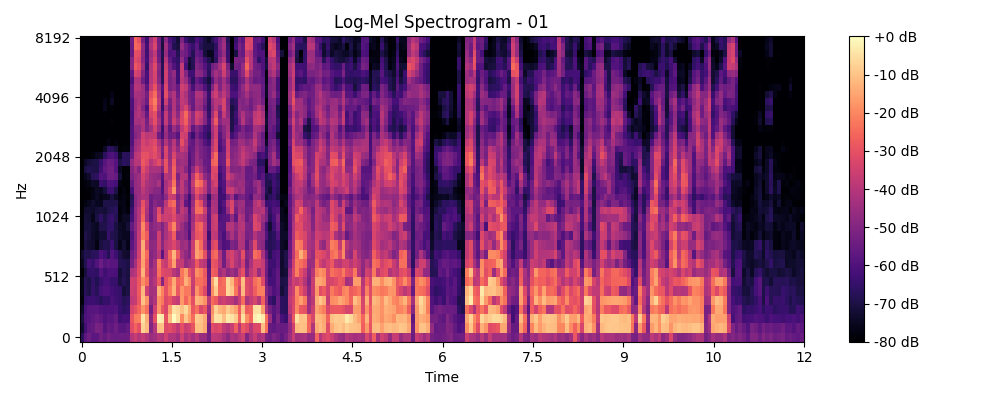}}
%%%call your figure name in the place "figurename.eps"
\caption{Example of spoken English exhibiting dense spectral energy across a wide frequency range, reflecting typical characteristics of natural speech.}
\label{fig_spec}
\end{figure}

\begin{itemize}

\item \textbf{Data projection:} All audio inputs were resampled to a common sampling rate (16 kHz) and amplitude-normalized to minimize variation due to recording conditions \citep*{ko2015audio,tawakuli2024preprocessing}. This standard preprocessing step ensures comparability across input types and reduces signal artifacts due to inconsistent recording. Each file was segmented into fixed-duration windows (12 seconds), converted into log-mel spectrograms \citep{xu2022differential}, and flattened into one-dimensional vectors. The log-mel features are commonly used to represent time-frequency structure in audio analysis. Figure 2 presents a representative log-mel spectrogram from the human input category.

\item \textbf{Tokenization:} 
 The flattened spectral vectors were then used as the basis for unsupervised clustering \citep[e.g., K-means;][]{liu2020model}, producing sequences of discrete symbolic tokens.
These symbolic tokens aim to capture structural regularities within the spectro-temporal patterns of the input.
Clustering was applied along the time axis of the log-Mel spectrogram, preserving sequential structure in the symbolic representation. 
This ensures that pattern transitions over time are retained in the input prompt to the model.
This clustering-based tokenization aligns with prior work in self-supervised speech representation learning, where discrete units are used as training targets for predictive modeling \citep{liu2020model, hsu2021hubert}.

\item \textbf{LLM provocation:} The resulting token sequences are passed as prompts to a pre-trained language model (e.g., GPT-2), without conditioning or instruction. We observe whether the model generates coherent output with sampling disabled and response length constrained. This setup allows consistent evaluation across inputs while minimizing variability due to randomness.

\item \textbf{Semantic response measurement:} We used the metric SIP to quantify the model’s generative behavior that integrates four subcomponents:
\begin{enumerate}
\item \textbf{Token-level entropy} ($H_{\text{token}}$), reflecting uncertainty in the output. This is computed as the average Shannon entropy over the model’s token probability distribution \citep{shannon1948}, measuring how confidently the model selects each token. The Shannon entropy is computed as:
\[
H_{\text{token}} = - \frac{1}{N} \sum_{t=1}^{N} \sum_{i=1}^{V} p_{t,i} \log p_{t,i}
\]

where \(N\) is the number of tokens in the output, \(V\) is the vocabulary size, and \(p_{t,i}\) is the predicted probability of token \(i\) at position \(t\).

\item \textbf{Syntax coherence score}, following standard practice in evaluating language model outputs, we estimate syntax coherence using the inverse token-level cross-entropy, computed via a pretrained autoregressive model \citep[GPT-2; ][]{Radford2019GPT2}. Lower loss indicates higher syntactic fluency and internal consistency.
We compute:
\[
\text{Syntax}_{\text{score}} = \frac{1}{\mathcal{L}_{\text{CE}} + \varepsilon}
\]
where \(\mathcal{L}_{\text{CE}}\) is the token-level cross-entropy loss, and \(\varepsilon = 10^{-5}\) avoids division by zero.

\item \textbf{Compression gain}, We estimate compression gain as a measure of structural regularity, using the relative reduction in UTF-8 byte length after applying zlib compression. Higher compression gain indicates greater internal redundancy or patterning within the model’s output, reflecting a lower-entropy, more structured sequence \citep{deletang2023compression}.

\[
\text{Compression Gain} = 1 - \frac{L_{\text{compressed}}}{L_{\text{original}}}
\]

\noindent
where $L_{\text{original}}$ is the byte length of the uncompressed text (encoded in UTF-8), and $L_{\text{compressed}}$ is the byte length of the same text after zlib compression using the LZ77 algorithm \citep{deletang2023compression}.

\item \textbf{Repetition penalty}, penalizing excessive token-level redundancy. This metric quantifies how often the same tokens appear multiple times in the generated output, which may indicate degenerative looping behavior or a lack of structural variety. Repetition is measured as the proportion of tokens that occur more than once in the generated sequence:

\[
\text{Repetition Penalty} = \frac{N_{\text{repeat}}}{N_{\text{total}}}
\]

where \( N_{\text{repeat}} \) is the number of unique tokens that appear more than once, and \( N_{\text{total}} \) is the total number of tokens in the output. Higher values suggest reduced generative diversity and are therefore penalized in the final SIP score.

\end{enumerate}

These components were selected to capture different dimensions of generative structure, including uncertainty, syntactic form, compressibility, and redundancy. 
Together, they contribute to a composite score designed to quantify linguistic reactivity.

The SIP is computed as:
\[
\text{SIP} = \alpha \cdot (1 - H_{\text{token}}) + \beta \cdot \text{Syntax}_{\text{score}} + \gamma \cdot \text{Compression}_{\text{gain}} - \delta \cdot \text{Repetition}_{\text{penalty}}
\]

The coefficients $\alpha$, $\beta$, $\gamma$, and $\delta$ represent the relative weight of each component in the final score. In this study, we set $\alpha = 2.0$, $\beta = 1.5$, $\gamma = 1.0$, and $\delta = 0.5$ to give higher priority to entropy and syntactic coherence, as these components showed stronger effects in smaller models. Compression gain and repetition penalty were also included with smaller weights to preserve structural variation.

\item \textbf{Baseline comparison:} We compared SIP scores across the four input types. SIP is designed to reflect the potential of each input to trigger language-like output, regardless of its semantic content. This allows us to detect structural reactivity in the data without assuming any prior symbolic intent or communicative purpose.

\end{itemize}

\section{Results}
\subsection{Initial Demonstration}

Our SIP results for each input category are shown in Figure~3 and summarized in Table~1, allowing for comparison of within-class variation. 
These visualizations provide a practical demonstration of the metric’s ability to differentiate structural reactivity across input types. 
Inputs with internal organization, such as birdsong and whale vocalizations, tend to trigger stronger responses than white noise. 
This suggests that the model is responding to structural patterns in the data, rather than to any semantic content.

Interestingly, this sensitivity is not limited to broad input categories. Even within the same class, such as humpback whale vocalizations, SIP scores can vary significantly depending on the specific recording (Table~2). One group of whale clips showed only moderate reactivity, while another gave high SIP values across all measures. This suggests that the method responds to differences in internal structure, not to species type or the presence of meaning. In this view, SIP works more like a detector of pattern density than a classifier. Even sounds from the same animal can range from unstructured to highly evocative, depending on the internal organization of the data.

One surprising result is that human speech scored only slightly higher than white noise, and much lower than whale and bird vocalizations. This suggests that the model does not treat human speech as structurally unique. Since the LLM receives all inputs as symbolic sequences, differences in SIP are more likely to arise from data structure than from training-related priors. From the model’s perspective, human language may appear less internally consistent than other input types. In some cases, it even resembles noise more than structured patterns like birdsong. Rather than reacting to semantic content, the model responds to properties such as entropy and compressibility, which do not always match human intuitions about meaning. In this sense, language may not be the clearest sign of intelligence, but just one form of structure among others. 

The relatively lower SIP scores for human speech may reflect structural properties in the tokenized representation, rather than any learned semantic bias in the model.

This reinforces a key idea behind the method: it is not designed to favor human communication, but to detect structure wherever it appears, even if it comes from unfamiliar or non-linguistic sources.

\begin{figure}[!h]
\centerline{\includegraphics[width=5.5in]{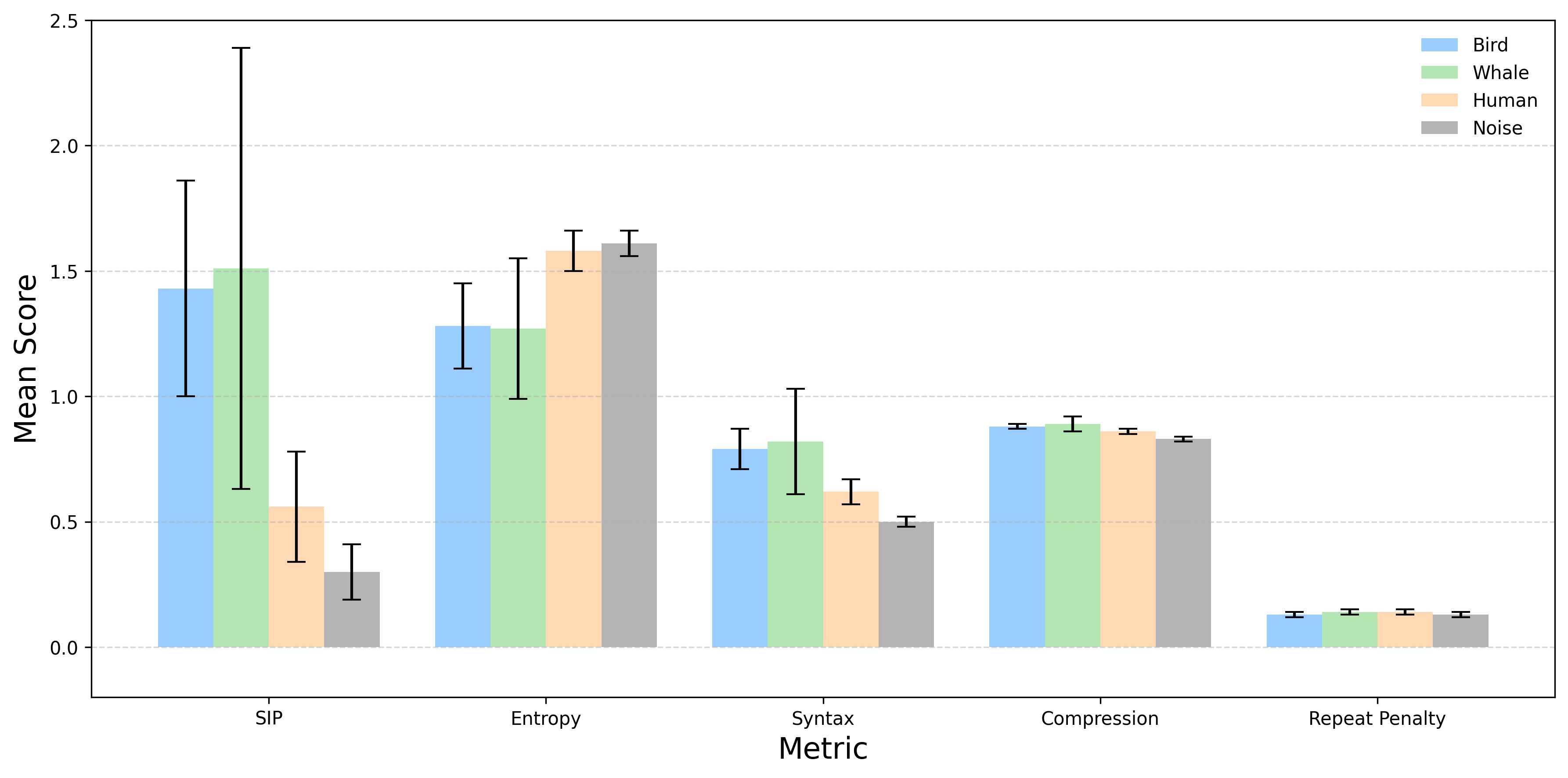}}
%%%call your figure name in the place "figurename.eps"
\caption{Mean scores for SIP and its components (entropy, syntax, compression, and repetition) across four input types.}
\label{fig_sim}
\end{figure}

\begin{table}
\centering
\small
\tabcolsep=6pt
\caption{Semantic Induction Performance (SIP)\label{table_example}}
\begin{tabular}{llllll}
\hline
         & SIP             & Entropy         & Syntax          & Compression     & Repeat Penalty \\
\hline
Noise    & $0.30 \pm 0.11$ & $1.61 \pm 0.05$ & $0.50 \pm 0.02$ & $0.83 \pm 0.01$ & $0.13 \pm 0.01$ \\
Human    & $0.56 \pm 0.22$ & $1.58 \pm 0.08$ & $0.62 \pm 0.05$ & $0.86 \pm 0.01$ & $0.14 \pm 0.01$ \\
Whale    & $1.51 \pm 0.88$ & $1.27 \pm 0.28$ & $0.82 \pm 0.21$ & $0.89 \pm 0.03$ & $0.14 \pm 0.01$ \\
Bird     & $1.43 \pm 0.43$ & $1.28 \pm 0.17$ & $0.79 \pm 0.08$ & $0.88 \pm 0.01$ & $0.13 \pm 0.01$ \\
\hline
\end{tabular}
\vspace{1ex}
\begin{flushleft}
\textit{Notes:} Related indicators across species and data types. Values are mean $\pm$ standard deviation.
\end{flushleft}
\end{table}

\begin{table}
\centering
\small
\tabcolsep=6pt
\caption{SIP Scores for Whale Segments by Recording Source\label{table_example}}
\begin{tabular}{llllll}
\textbf{Category} & \textbf{SIP} & \textbf{Entropy} & \textbf{Syntax} & \textbf{Compression} & \textbf{Repeat Penalty} \\
\hline
CI05 (Chunks 1--5) & $0.68 \pm 0.16$ & $1.54 \pm 0.06$ & $0.64 \pm 0.04$ & $0.86 \pm 0.01$ & $0.13 \pm 0.003$ \\
CI02 (Chunks 6--10) & $2.34 \pm 0.36$ & $1.01 \pm 0.09$ & $1.01 \pm 0.12$ & $0.91 \pm 0.01$ & $0.14 \pm 0.01$ \\
\hline
\end{tabular}
\vspace{1ex}
\begin{flushleft}
\textit{Notes:} Each row summarizes the average SIP and its components for whale segments from two distinct recordings (CI05 and CI02). Values reflect the mean $\pm$ standard deviation across five 12-second segments per source.
\end{flushleft}
\end{table}

\subsection{Interpretation}

Although we use only short audio segments and a single model, the results show that structured input can trigger language-like responses in a generative system. These inputs do not need to carry recognizable meaning or follow linguistic rules. More broadly, semantic triggering does not depend on prior exposure to language. Generative models can be tested for reactivity without assuming any specific communicative form.

From a SETI perspective, this points to a possible shift in detection strategy. If generative models can respond to structured input without replying on semantic content, then it may be more productive to focus on data that is rich in internal patterns, even if it does not resemble language. Detection efforts could benefit from models that react to structural regularities, rather than relying on symbolic formats or assumed communicative intent.

Moreover, early-stage civilizations may rely on strong signals to overcome distance and background noise. But this approach is energy-intensive and may not last long. In contrast, more advanced systems might use low-power, long-duration transmissions that hide structure within data that appears noise-like. Such data may persist longer and remain harder to detect, unless the receiver is sensitive to internal structure.

Although these patterns may not have conventional markers, they could still be preserved in archival data or transmitted in formats that last over time. Their low-energy, noise-like form is often ignored by traditional filters. But when reanalyzed with a structure-sensitivity approach, they may still trigger structured responses. This makes them more likely to be detected, especially if the receiver is tuned to hidden structure instead of semantic content. This opens the possibility of finding overlooked patterns in data that were previously discarded as noise.

The Drake equation’s \citep{drake1961} most uncertain variable, \(L\), representing the lifetime of a technological civilization, can be refined. Instead of considering only the civilization’s duration (\(L_0\)), we also include the persistence of its structural imprint in recorded data (\(T_p\)). This refers to the time that patterns remain detectable through structural analysis, even after transmission ends. With this, the effective detection window becomes:

\[
L_{\text{effective}} = L_0 \cdot T_p
\]

This idea shows that detectable patterns may outlast the civilizations that produced them, shifting the focus from real-time contact to searching through archives. Detectability depends not only on how long a civilization exists, but also on how long its emissions retain a structure that can still be found in data. Traditional detection methods rely on fixed templates and often ignore data that lacks expected features. In contrast, the STDP framework enables re-examination of data based on model-driven reactivity. Patterns that were previously overlooked may become detectable through this kind of reanalysis. This may support a shift from real-time monitoring to long-term data mining in SETI efforts.

\section{Discussion}
\subsection{Complementary Strategies for SETI and Beyond}
The STDP framework offers a different way to approach detection in SETI. It treats detection not as a decoding task, but as a test of reactivity to structural patterns. Instead of assuming that extraterrestrial data is meant to be understood, this method asks whether structure alone can trigger generative responses in models trained on human language. This view leaves the possibility that advanced civilizations may prefer to send patterns that are only detectable by receivers with compatible capabilities.

In this view, language models act as detectors of structural patterns. They respond to patterns in the input, even without any recognizable content. The STDP method adds a new way to analyze SETI data, based on how sensitive a model is to hidden structure in the data. We call this property \textit{semantic evocativity}, defined as the tendency of a generative model to produce structured language when given nonlinguistic or unlabeled input. Our results show that this reactivity can be measured and used to detect structure without relying on labels or decoding. 

This approach opens new questions about how models perceive structure, and how such sensitivity might be trained or tuned in future systems. One possibility is to develop lightweight structure-sensitive language models that can identify data with high potential to trigger generative behavior. Noise-like input streams could be tested for signs of internal patterning, such as more regular sentence structure, lower output entropy, or repeated use of related words. If these effects are consistent across model types and input sources, they would further support the use of generative reactivity as a viable filtering mechanism.

Our approach connects to ongoing research in machine learning, cognition, and bioacoustics. We refer to this direction as \textit{semantic neural evocativity}, a research area focused on how nonlinguistic input can lead to structured language output in artificial systems. This perspective shifts the focus from interpreting content to detecting structural responses.

From a cognitive standpoint, the ability to produce structured output in response to ambiguous or unlabeled input may reflect a threshold of abstraction. 
Some researchers have linked this threshold to concepts such as intentionality or synthetic awareness. In language modeling, this opens up a way to use LLMs not only for generation, but also to study how different conditions influence the emergence and quality of language-like behaviors \citep{holtzman2020degeneration,wei2022emergent}.

In animal communication studies, this method avoids the problem of cross-species translation by asking a simpler question: does the system produce structured output when exposed to input that contains patterns? In SETI, the same approach introduces SIP as an additional measure. Instead of evaluating data by its information content, it is assessed by its ability to trigger language-like responses.

\subsection{Practical Considerations for Real SETI Data}
To apply the STDP framework to real-world SETI radio data, such as baseband recordings or filterbank outputs, a number of preprocessing steps are required. 
These steps are nontrivial and introduce several practical considerations:

\begin{itemize}
\item \textbf{Discretization of continuous data:}
Discrete symbolic conversion is required to map continuous radio signals into sequences interpretable by language models. While this transformation is necessary, it risks introducing structural artifacts or obscuring patterns that may be present. Since STDP aims to detect such patterns through model response, care must be taken to ensure that any observed behavior is not an artifact of the preprocessing pipeline.

\item \textbf{Spectral leakage and calibration artifacts:} 
Instrumental features such as electronic harmonics, side lobes, or digitization effects may produce structured patterns in the spectral domain. These artifacts can lead to false-positive SIP activations by producing patterns that resemble structured input. Such systematics are well documented in SETI pipelines \citep{enriquez2017breakthrough, siemion2012fly}, and must be accounted for when interpreting reactivity in real data.

\end{itemize}

Addressing these considerations will be essential for future work aiming to integrate STDP into real-time SETI pipelines or large-scale archival analysis. 
In particular, validating SIP scores against known instrumental features, entropy baselines, or cross-modal consistency may help distinguish true data-driven responses from artifacts introduced by structural noise.

\subsection{Future Directions}

This work opens several directions for further investigation. One is the application of the SIP metric to public SETI archives, including baseband and filterbank datasets. For instance, large-scale radio surveys such as \citet{siemion2012fly} produce extensive background segments that are discarded due to lack of narrowband or transient features. The STDP framework could be applied to such archival datasets to evaluate whether any discarded segments trigger structured model responses, which may reflect latent structure in the data. Structural reactivity analysis could thus uncover structure that escapes conventional energy-based or periodicity-based filters, and complement existing anomaly-focused surveys.

A second direction is to establish SIP baselines across different stellar classes, target types, or regions of the sky. By computing reactivity distributions from archival data, it may be possible to identify outlier segments with unexpectedly high SIP scores. These anomalies could then be further investigated for interference, signal artifacts, or unexplained structure, with generative reactivity serving as a complementary detection dimension.

Third, this approach suggests a speculative but testable extension: communication synthesis guided by reactivity criteria. Structured inputs could be designed to resemble background noise while still containing enough internal organization to trigger responses in generative systems. To test for cross-system sensitivity to structure, their effects could be evaluated across different types of receivers, including humans, animals, and language models with varying training constraints.

This line of investigation points to a broader idea we call "cosmic linguistic seeding": a communication strategy based not on content delivery, but on the triggering of symbolic behavior in structurally responsive systems. The concept is related to previous proposals for minimum-energy interstellar communication \citep{messerschmitt2013interstellar}, but shifts the optimization goal from sending as much information as possible to triggering structured responses in systems that are ready to detect them. While speculative, this framework offers a concrete path toward implementing reactivity-based SETI, and may extend to artificial or non-human observers beyond Earth.

In addition, the framework offers new opportunities for participatory science. We propose that public computing platforms, such as SETI@home \citep{anderson2002seti}, could be adapted for structure-based data evaluation. In contrast to traditional SETI workflows that often rely on high-throughput spectral analysis (e.g., Fourier transforms across wide frequency ranges), this approach employs lightweight language models that can be executed locally with minimal computing resources. These models would process unlabelled data segments and compute SIP scores, enabling rapid screening of noise-like input. This setup allows broader participation without requiring users to decode signals or make semantic judgments, and may offer a new foundation for distributed, structure-sensitive astrobiological detection.

\section{Conclusion}

Conventional SETI searches often focus on signal amplitude, energy distribution, or frequency-domain anomalies. This work explores a different form of detectability that we call semantic potential. By this, we mean a model’s tendency to produce structured language when exposed to unlabeled input. Even if the origin or content of data is unknown, it may still provoke a measurable reaction.

To our knowledge, no previous study has systematically examined how entropy anomalies in archival SETI data relate to the way language models respond. We offer a framework that treats all inputs as unstructured and evaluates whether any segment leads to language-like output. The method does not try to decode signals. Instead, it tests whether structure emerges in how the model responds. These reactions might reveal patterns that conventional filters would miss.

This approach also offers a different way to think about active SETI. Rather than sending messages to be understood, an advanced civilization might send data meant to activate symbolic behavior in whoever receives them. The goal is not to instruct, but to provoke. We refer to this as cosmic linguistic seeding.

This work is not intended as a comprehensive classification of structural responsiveness across species, but as a proof-of-concept for the STDP framework and the SIP metric. While future studies may pursue broader validation, our goal here is to demonstrate that generative linguistic reactivity is both measurable and sensitive to structural patterns. More broadly, this work asks whether detectability must rely on embedded meaning, or whether structure alone can provoke linguistic behavior. Perhaps the most important signal is not a message, but the moment when noise begins to trigger linguistic behaviors.

\section{Acknowledgement}
P.-C.Y. is supported by the Taiwan Astronomical Research Alliance (TARA) and ASIAA, with funding from the National Science and Technology Council (NSTC 113-2740-M-008-005). TARA is committed to advancing astronomy in Taiwan and paving the way for the establishment of a national observatory. We thank the contributors of the Xeno-Canto community for providing open access to curated birdsong recordings, which enabled structured analysis of non-human vocal data. The specific Phylloscopus trochilus sample used in this study was recorded by Jens Kirkeby (XC763801). The data is used under the Creative Commons BY-NC-SA 4.0 license. We thank the NOAA Office of National Marine Sanctuaries and the U.S. Navy for providing open-access acoustic recordings through the SanctSound project, which were used for whale vocalization analysis in this study. Human speech data were obtained from LibriVox, a volunteer-driven platform providing public domain audiobook recordings (\url{https://librivox.org}). Recordings are in the public domain in the United States; copyright status may vary by country.

\end{document}